# Chiral effects at the metal center in Fe(III) spin crossover coordination salts.

M. Zaid Zaz,[1] Wai Kiat Chin,[1] Gauthami Viswan,[1] Arjun Subedi,[1] Esha Mishra,[1,2] Kayleigh A. McElveen,[3] Binny Tamang,[3] David Shapiro,[4] Alpha T N'Diaye,[4] Rebecca Y Lai[3] and Peter A Dowben[1]

[1]Department of Physics and Astronomy, Jorgensen Hall, University of Nebraska-Lincoln, Lincoln, NE 68588-0299, USA.
[2]Department of Physics, Berry College, 2277 Martha Berry Hwy. NW., Mount Berry, GA 30149, USA.
[3]Department of Chemistry, Hamilton Hall, University of Nebraska-Lincoln, Lincoln, NE, 68588-0304, USA.
[4]Advanced Light Source, Lawrence Berkeley National Laboratory, Berkeley, CA 94720, USA

E-mail: pdowben@unl.edu

**Abstract**

Evidence of chirality was observed at the Fe metal center in Fe(III) spin crossover coordination salts [Fe(qsal)$_2$][ Ni(dmit)$_2$] and [Fe(qsal)$_2$](TCNQ)$_2$ from X-ray absorption spectroscopy at the Fe 2p3/2 core threshold. This indicates the formation of chiral domains that influence the octahedral coordination on the Fe core.

## 1. Section heading

Chiral behaviour has been observed in several spin crossover molecular complexes[1–18] and is generally expected. Chirality in spin crossover systems is often a result of inversion symmetry breaking due to the presence of asymmetric ligands. Chirality can also arise due to distorted octahedral coordination resulting from bulky functional groups attached to the ligands or also due to the presence of bulky counterions[10,14,19–21] or, in principle, the molecular packing. Yet demonstrations of chirality in visible[1–10] and infra-red[8] spectra do little to address the effect of chirality on the metal center weighted molecular orbitals. So, while chirality is common, we ask here whether the chirality from molecular packing and ligand choice influence the octahedrally coordinated iron. X-ray circular dichroism or X-ray natural circular dichroism has been observed in the X-ray from both metal centered molecular complexes,[22] or an organic system, typically containing a somewhat larger Z atom, like Cl in chlorophenylethanol or chlorohexahelicene.[23] This X-ray circular dichroism (XCD) or X-ray natural circular dichroism (XNCD) is distinct from X-ray magnetic circular dichroism (XMCD), as a magnetic moment is not required for XCD/XNCD, just the loss of inversion symmetry.[24]

In this paper, we present evidence of chirality in Fe(III) spin crossover coordination salts, namely [Fe(qsal)$_2$][Ni(dmit)$_2$], [Fe(qsal)$_2$](TCNQ)$_2$ where, (qsalH = N(8quinolyl)salicylaldimine), (dmit$_2^-$ = 1,3-dithiol-2-thione-4,5-dithiolato) and (TCNQ= 7,7,8,8-tetracyanoquinodimethane) from spatially resolved X-ray absorption spectroscopy using circularly polarized X-rays. These chiral salts are similar to spin crossover chiral salt complexes that feature large conductance, where FeIII(qsal)$_2$Ni(dmit)$_2$]$_3$.CH3CN.H2O has a resistance less than 1 ohm.cm,[25] as does [Fe(III)(3-OMe-sal$_2$trien)][Ni(dmit)$_2$][26] and [Fe(III)(sal2trien)](TCNQ)2·CH3OH,[27] while Fe(sal$_2$trien)[Ni(dmit)$_2$]$_3$ has a resistance less than 3 Ohm.cm,[28] all when in the high spin state at elevated temperatures. These low resistivities are atypical of spin crossover complexes that have no counter ion,[29-31] although a counter ion like TCNQ does not ensure this exceptional conductivity.[31-34]

The Fe(III) coordination complex, [Fe(qsal)$_2$]Cl where (qsalH = N(8quinolyl)salicylaldimine), was synthesized in accordance with a protocol reported elsewhere.[35] The procedure for the synthesis of [Fe(qsal)$_2$](TCNQ)$_2$ followed the procedure for the synthesis of [Fe(qsal)$_2$]Cl, followed by a process detailed elsewhere.[27,34] [Fe(qsal)$_2$][ Ni(dmit)$_2$] was also synthesized according to a protocol reported earlier[36] (see electronic supplementary material for details on confirmation of successful synthesis of all the three complexes). The Fe to Ni ratio for [Fe(qsal)$_2$][ Ni(dmit)$_2$] was confirmed by X-ray diffraction (XRD) and energy dispersive analysis of emitted X-rays (EDAX). While the Fe to Ni ratio is 1:1 from EDAX, consistent the molecular stoichiometry of [Fe(qsal)$_2$][Ni(dmit)$_2$], X-ray photoemission (XPS) studies, from the Fe and Ni 2p core levels, indicate that the Fe to Ni ratio is 1:2. This suggests that the surface terminates in Ni(dmit)$_2$, as discussed elsewhere.[37] Consistent with a Ni(dmit)$_2$ surface termination, the surface has a strong surface to bulk core level shift in the Ni 2p core level XPS spectra, resulting from final state effects.[37]

The magnetometry measurements of DC magnetic susceptibility times temperature were carried out in DynaCool SQUID magnetometer from Quantum Design at the Nebraska Center for Materials and Nanoscience. Magnetometry measurements (Figure 1) performed on [Fe(qsal)$_2$X], X = Ni(dmit)$_2$, Cl, or TCNQ$_2$, match well with what is reported in the literature on [Fe(qsal)$_2$]Cl[35] and [Fe(qsal)$_2$][Ni(dmit)$_2$].[25,36] The magnetometry of [Fe(qsal)$_2$](TCNQ)$_2$ reveals a gradual spin transition with increasing temperature much like [Fe(qsal)$_2$][ Ni(dmit)$_2$] and indicates that at room temperature, these three Fe(III) complexes are in the high spin state or nearly so.

Spatially resolved X-ray absorption spectroscopy experiments, using left and right circularly polarized light, were performed on the COSMIC beamline (Beamline 7.0.1.2) at the Advanced light source in Lawrence Berkeley national laboratories. The incident beam was tuned to the Fe L3 edge (Fe 2p$_{3/2}$) at 707 eV, using a small spot size (50 nm) to be able to perform the XAS, in the transmission mode, on a single Fe(qsal)$_2$X chiral crystallite domain, X = Ni(dmit)$_2$, Cl, or TCNQ$_2$. This resembles the spatially resolved element-specific X-ray natural circular dichroism studies of multiferroic crystals.[38] The relative local Ni and Fe moments were investigated with x-ray absorption (XAS) and X-ray magnetic circular dichroism (XMCD) spectroscopies at beamline 6.3.1 (Advanced Light Source, Berkeley, CA), using total electron yield (TEY) and circularly polarized x-rays with an estimated degree of polarization of 0.66. The XMCD spectra were obtained with an applied field of 1.8 T. The inverse photoemission (IPES) experiments were acquired in the isochromatic mode with a commercial high resolution electron gun (Kimball Physics) combined with channeltron-based photon detector (OmniVac).

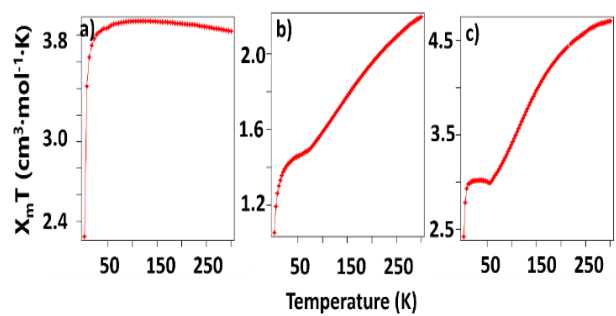

**Fig. 1** The Product of molar magnetic susceptibility and temperature of: [Fe(qsal)$_2$]Cl, b) [Fe(qsal)$_2$][Ni(dmit)$_2$], c) [Fe(qsal)$_2$](TCNQ)$_2$, plotted against temperature.

Figure 2 shows the X-ray absorption spectra at the Fe L2 (2p$_{1/2}$) and L3 (2p$_{3/2}$) edges, for single chiral domains of Fe(qsal)$_2$X, X = Ni(dmit)$_2$, Cl, and TCNQ$_2$, obtained using right and left circularly polarized X-ray radiation in the absence of any applied external field. For [Fe(qsal)$_2$](TCNQ)$_2$, right circularly polarized X-rays excite the Fe 2p$_{3/2}$ core electrons to populate both the unoccupied t$_{2g}$ and e$_g$ states, resulting in two distinct peaks at the Fe L3 (2p$_{3/2}$) edge, while as for left circularly polarized X-rays core excitation favours the population of the unoccupied t$_{2g}$ states at lower photon energies. This indicates the existence of a circular polarization dependent selection rule. Similar behaviour is observed for [Fe(qsal)$_2$][ Ni(dmit)$_2$], though a little less dramatic. However, for [Fe(qsal)$_2$]Cl, we see that the absorption spectra for both right and left circularly polarized X-rays are similar, although still not identical.

Figure 3 shows that there is a small X-ray magnetic circular dichroism signal (4%) at the Fe 2p core edges, for [Fe(qsal)$_2$][ Ni(dmit)$_2$] under 1.8 T applied magnetic field. This XMCD signal resides largely under the Fe 2p3/2 core to bound absorption, indicative that the Fe spin moment dominates. The Fe moment from the Fe(qsal)$_2$ moiety is antiparallel with the moment from the Ni(dmit)$_2$ moiety, as the XMCD response is opposite in sign with respect to the applied magnetic field, as seen in Figure 4. The XMCD spectra of Figure 3 do not resemble Figure 2. Thus, not only is the Ni antiferromagnetically coupled to the Fe metal center, but the X-ray chiral dichroism signal of [Fe(qsal)$_2$][ Ni(dmit)$_2$], shown in Figure 2, is very distinct from the magnetic circular dichroism spectra of Figure 3.

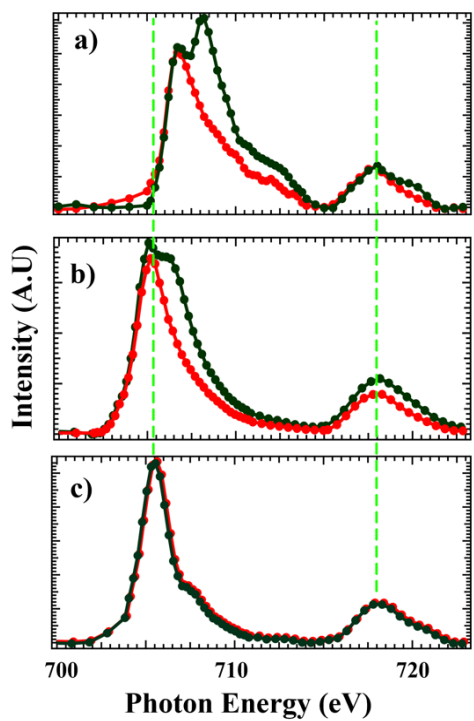

**Fig. 2** The Fe L2 ($2p_{1/2}$) and L3 ($2p_{3/2}$) edge X-ray absorption spectra of a) [Fe(qsal)$_2$](TCNQ)$_2$, b) [Fe(qsal)$_2$Ni(dmit)$_2$], c) [Fe(qsal)$_2$]Cl, obtained using right circularly polarized (blue) and left circularly polarized (red) X-ray radiation.

Because inverse photoemission is very surface sensitive and because the surface of the [Fe(qsal)$_2$][Ni(dmit)$_2$] thin film terminates in Ni(dmit)$_2$,[37] as determined by X-ray photoemission, the inverse photoemission is representative mostly of the 1,3-dithiol-2-thione-4,5-dithiolato ligand (dmit) unoccupied density of states. This unoccupied ligand (dmit) density of states overlaps the region where the XMCD signal is the greatest, at photoenergies well above the Ni $2p_{3/2}$ core threshold. Thus, a substantial fraction of the moment on the Ni(dmit)$_2$ moiety is seen to reside on the ligand itself and implies that antiferromagnetic coupling between the Ni and Fe is mediated by the ligand.

The light circular polarization dependence of the X-ray absorption of [Fe(qsal)$_2$](TCNQ)$_2$ and [Fe(qsal)$_2$][Ni(dmit)$_2$] (Figure 2a and 2b) can either be a result of magnetic circular dichroism (XMCD) or X-ray chiral effects, i.e. X-ray circular dichroism (XCD) or X-ray natural circular dichroism (XNCD).[24] As the circular dichroism results, of Figure 2, were obtained in the absence of an applied magnetic field, and do not resemble the X-ray magnetic circular dichroism of Figure 3, or more correctly, the X-ray magnetic circular dichroism resembles the sum, not either of the X-ray absorption spectra taken as a function of circular polarization. The X-ray absorption dependence on light circular polarization, observed at the Fe $2p_{3/2}$ core, is thus an X-ray chiral circular dichroism effect implying loss of inversion symmetry, not an XMCD signal resulting from a net magnetic moment.[24]

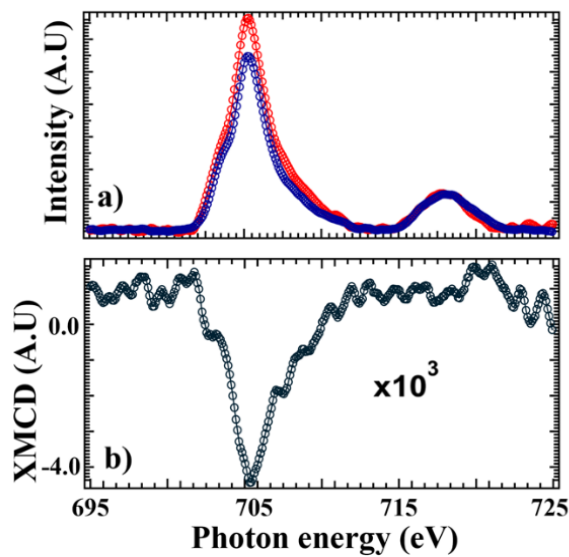

**Fig. 3** The Fe L2 ($2p_{1/2}$) and L3 ($2p_{3/2}$) edge X-ray absorption spectra of [Fe(qsal)2Ni(dmit)2] at the under opposite 1.8 T applied magnetic fields (a) and the resulting magnetic circular dichroism signal (b).

The chiral texture affects the Fe weighted molecular orbitals. Chirality imparts spin-orbit coupling, due to the loss of inversion symmetry. The strong spin-orbit coupling that now appears within the Fe$2p_{3/2}$ envelope, is partially subtracted the Fe $2p_{3/2}$ to $2p_{1/2}$ spin-orbit splitting as the latter is diminished for [Fe(qsal)$_2$](TCNQ)$_2$ and [Fe(qsal)$_2$][Ni(dmit)$_2$] compared to [Fe(qsal)$_2$]Cl, as seen in Figure 2. It is noteworthy that these chiral spin crossover thin films possess multiple domains, that is to say that the circular polarization dependence of the X-ray irradiation occurs, for both left and right circularly polarized X-rays, becomes negligible if a large number of domains are included. The X-ray absorption differences, as a function of incident X-ray circular polarization, varies in spatially disjoint regions, indicating that chiral crystallite domains of both molecular enantiomers exist.

The XAS Fe $2p_{3/2}$ to Fe $2p_{1/2}$ spin orbit splitting is significantly smaller (approximately 0.6 to 2.4 eV) for [Fe(qsal)$_2$](TCNQ)$_2$ that for either [Fe(qsal)$_2$][Ni(dmit)$_2$] (approximately 0.6 eV) or [Fe(qsal)$_2$]Cl (approximately 2.4 eV), as seen in Figure 2. The Fe $2p_{3/2}$ to Fe $2p_{1/2}$ spin orbit splitting, seen in XAS for [Fe(qsal)$_2$]Cl (Figure 2c), is roughly 12.7 eV, close to the value expect for the Fe 2 core (13 eV). The perturbation to the Fe 2p spin-orbit splitting, seen in XAS for [Fe(qsal)$_2$](TCNQ)$_2$ (Figure 2a) and to a much smaller extent for [Fe(qsal)$_2$][Ni(dmit)$_2$] is unlikely to be an initial state effect. The initial state spin-orbit splitting of a core level should be dominated by the atomic spin-orbit splitting, which should be insensitive to the ligand field. The fact that the perturbations to the Fe 2p spin-orbit splitting [Fe(qsal)]$(TCNQ)_2$ (Figure 2a) and [Fe(qsal)$_2$][ Ni(dmit)$_2$] are largest in the XAS final state where the chiral effects are greatest is noteworthy, as loss of inversion symmetry does impart spin-orbit coupling.

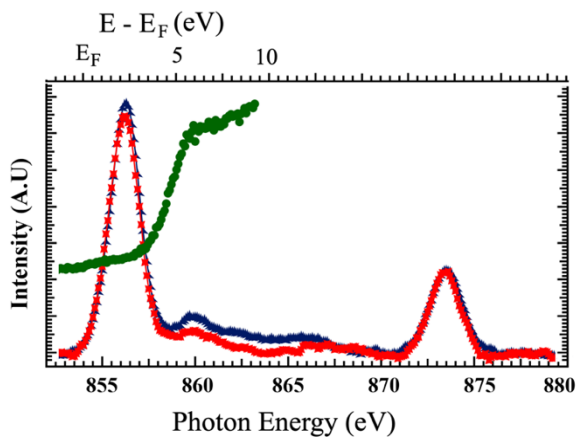

Fig. 4 The Ni L2 (2p1/2) and L3 (2p3/2) edge X-ray absorption spectra of [Fe(qsal)$_2$Ni(dmit)$_2$] at the under opposite 1.8 T applied magnetic fields (blue and black) and the resulting inverse photoemission spectra aligned at the 2p3/2 edge, indicating the unoccupied density of states above the chemical potential $E_F$, as indicated by the scale bar at the top.

In summary, we have shown the presence of a strong chiral texture in Fe(III) spin crossover coordination salts with heavy counterions. The Fe core is sufficiently perturbed within a single chiral domain crystallite to result in strong chiral effects at the Fe 2p3/2 core threshold in circularly polarized light dependent X-ray absorption spectroscopy of the Fe(III) spin crossover coordination salts [Fe(qsal)$_2$][ Ni(dmit)$_2$] and [Fe(qsal)$_2$](TCNQ)$_2$ but not [Fe(qsal)$_2$]Cl. These X-ray chiral effects observed for an individual crystallite domain of [Fe(qsal)$_2$Ni(dmit)$_2$] are distinct from X-ray magnetic dichroism effects and thus constitute an X-ray natural circular dichoism distinct from XMCD. Magnetic dichroism was also observed and indicate that Fe center is antiferromagnetically aligned with Ni in [Fe(qsal)$_2$Ni(dmit)$_2$].

## Acknowledgements

This work was supported by the National Science Foundation (NSF) through the NSF-DMR-EPM 2317464 (MZZ, WKC, GV, EM and PAD), and EPSCoR RII Track-1: Emergent Quantum Materials and Technologies (EQUATE) (KAM, RL and BT). This research used resources of the Advanced Light Source, which is a DOE Office of Science User Facility under contract no. DE-AC02-05CH11231.